\documentclass[manuscript]{aastex}

\shorttitle{P/2011 S1 (Gibbs) in PS1}
\shortauthors{Lin et al.}

\begin{document}


\title{Pan-STARRS 1 observations of the unusual active Centaur P/2011 S1(Gibbs)}


\author{H. W. Lin\altaffilmark{1}}
\affil{Institute of Astronomy, National Central University, Taoyuan 32001, Taiwan}
\email{edlin@gm.astro.ncu.edu.tw}

\author{Y. T. Chen\altaffilmark{2}}
\affil{Institute of Astronomy and Astrophysics, Academia Sinica, P. O. Box 23-141, Taipei 106, Taiwan}

\author{P. Lacerda\altaffilmark{3}}
\affil{Astrophysics Research Centre, School of Mathematics and Physics, QueenÕs University Belfast, Belfast BT7 1NN}

\author{W. H. Ip\altaffilmark{1}}
\affil{Institute of Astronomy, National Central University, Taiwan, 32001}

\author{M. Holman\altaffilmark{4} and P. Protopapas\altaffilmark{4}}
\affil{Harvard-Smithsonian Center for Astrophysics, 60 Garden Street, Cambridge, MA 02138, USA}

\author{W. P. Chen\altaffilmark{1}}
\affil{Institute of Astronomy, National Central University, Taiwan, 32001}

\and
\author{W. S. Burgett\altaffilmark{5}, 
K. C. Chambers\altaffilmark{5}, 
H. Flewelling\altaffilmark{5}, 
M. E. Huber\altaffilmark{5}, 
R. Jedicke\altaffilmark{5}, 
N. Kaiser\altaffilmark{5}, 
E. A. Magnier\altaffilmark{5}, 
N. Metcalfe\altaffilmark{5}, 
P. A. Price\altaffilmark{6}, 
}

\altaffiltext{1}{Institute of Astronomy, National Central University, Taoyuan 32001, Taiwan}
\altaffiltext{2}{Institute of Astronomy and Astrophysics, Academia Sinica, P. O. Box 23-141, Taipei 106, Taiwan}
\altaffiltext{3}{Astrophysics Research Centre, School of Mathematics and Physics, QueenÕs University Belfast, Belfast BT7 1NN}
\altaffiltext{4}{Center for Astrophysics, 60 Garden Street, Cambridge, MA 02138, USA}
\altaffiltext{5}{Institute for Astronomy, University of Hawaii, 2680 Woodlawn Drive, Honolulu HI 96822}
\altaffiltext{6}{Department of Astrophysical Sciences, Princeton University, Princeton, NJ 08544, USA}

\begin{abstract}

P/2011 S1 (Gibbs) is an outer solar system comet or active Centaur
with a similar orbit to that of the
famous 29P/Schwassmann-Wachmann 1. P/2011 S1 (Gibbs) has been observed by the Pan-STARRS 1 (PS1) sky survey
 from 2010 to 2012.  The resulting data allow us to perform multi-color studies of the nucleus and coma of the comet.
 Analysis of PS1 images reveals that P/2011 S1 (Gibbs) has a small
 nucleus $< 4$ km radius, with colors $g_{P1}-r_{P1} = 0.5 \pm 0.02$,
 $r_{P1}-i_{P1} = 0.12 \pm 0.02$ and $i_{P1}-z_{P1} = 0.46 \pm 0.03$.
 The comet remained active from 2010 to 2012, with a model-dependent
 mass-loss rate of $\sim100$ kg s$^{-1}$. The mass-loss rate per unit
 surface area of P/2011 S1 (Gibbs) is as high as that of
 29P/Schwassmann-Wachmann 1, making it one of the 
most active Centaurs. The mass-loss rate also varies with time from
$\sim 40$ kg s$^{-1}$ to 150 kg s$^{-1}$. Due to its rather circular
orbit, we propose that P/2011 S1 (Gibbs) has 29P/Schwassmann-Wachmann 1-like
outbursts that control the outgassing rate. The results indicate that
it may have a similar surface composition to that of
29P/Schwassmann-Wachmann 1. 

Our numerical simulations show that the future orbital evolution of
P/2011 S1 (Gibbs) is more similar to that of the main population of Centaurs
than to that of 29P/Schwassmann-Wachmann 1. The results also demonstrate that
P/2011 S1 (Gibbs) is dynamically unstable and can only remain near its
current orbit for roughly a thousand years. 
\end{abstract}


\section{Introduction}

The Centaurs are solar system bodies with orbits among four giant planets.
For this reason their orbits are unstable, their past and future
trajectories are typically chaotic, and their dynamical lifetimes are short \citep{tis03, hor04}.
Many theoretical investigations consider that this class of object is
the transitional population between the Kuiper belt objects and the
Jupiter-family comets (JFCs)\citep{fer94, lev97, tis03, eme05}. The
origin of Centaurs is still unclear, but widely accepted sources are
the Oort cloud and the scattered disk \citep{eme05}. 

The close relation with the JFCs suggests that some Centaurs may have
cometary-like activity. Indeed, the prototype of the Centaurs - (2060)
Chiron - has been shown to display cometary activity
\citep{mee90}. Several other ``active Centaurs" have been
identified \citep{jew09}. This kind of object prompted searches for
evidence of volatile materials. CN and CO have been detected in the
coma of (2060) Chiron \citep{bus91, wom97}.  Water ice has also been
reported on the surface of (2060) Chiron; its detectability is
correlated with the level of cometary activity. \citep{luu00, rom03}.  

29P/Schwassmann-Wachmann 1, which we refer to as 29P/SW1, is highly
active and is dynamically both a JFC (defined by Tisserand
parameter with respect to Jupiter, $T_{J}$,  $2 < T_{J} < 3$,
\cite{gla08}) and a Centaur (defined by semi-major axis and
perihelion, $a_J < q <a_N$ and $a_J < a < a_N$, where $a_J$ and $a_N$
are the semi-major axes of Jupiter and Neptune). Also a Centaur can not be in a 1:1 mean motion resonance with any planet \citep{jew09}.   
Its very circular orbit (eccentricity $\sim 0.04$), large perihelion
distance (5.72AU), and repeated outburst events set it apart from
other comets \citep{tri08, tri10}. CO has been detected in several
studies \citep{coc82, sen94, rea13} and is believed to be the source
of activity. The outbursts may relate to the nucleus rotation \citep{tri10}.   

In this work, we investigate a new active Centaur -- P/2011 S1 (Gibbs).
P/2011 S1 (Gibbs) was discovered by A. R. Gibbs on September 18, 2011 using the Mt. Lemmon 1.5-m reflector \citep{gib11}.
JPL classified this object as a Chiron-type comet, which is defined with
$T_J > 3$ and $a > a_J$ \citep{lev97}.  $T_J = 3.12$ for P/2011 S1 (Gibbs).
However, P/2011 S1 (Gibbs) has a small orbital eccentricity $\sim 0.2$,
which means it has a quite circular orbit, similar to 29P/SW1. 
(\citet{lac13} reported on another similar object, P/2010 TO20
LINEAR-Grauer (P/LG), as a possible mini 29P/SW1 with JPL
classification as JFC ($2 < T_{J} < 3)$.)

The Panoramic Survey
Telescope And Rapid Response System-1 (Pan-STARRS-1, PS1) has
serendipitously observed P/2011 S1 (Gibbs) several times, allowing us to
measure astrometry and multi-color photometry at a number of epochs.
These data enable comparison of the physical and orbital properties of
P/2011 S1 (Gibbs) to those of 29P/SW1 and (2060) Chiron.  This work is divided
into two main parts.  In the first we present the PS1 observations,
our photometry, and our analysis of the cometary activity of P/2011 S1
(Gibbs). In the second part we present the results of our numerical
integrations, comparing the orbital evolution of P/2011 S1
(Gibbs) with that of 29P/SW1 and (2060) Chiron. 

\section{Observations}

P/2011 S1 (Gibbs) was observed as a part of the PS1 survey. 
The PS1 telescope is a 1.8-m Ritchey-Chretien reflector located on Haleakala, Maui, 
which is equipped with a 1.4 gigapixel camera covering 7 square degree on the sky.
The PS1 survey has five different survey modes  \citep{kai10}. (1) The $3\pi$
Steradians survey, which repeatedly covers the $3\pi$ steradians of
sky visible from Haleakala and uses a photometric system that closely
approximates the SDSS filter system $g_{P1}$ (bandpass $\sim$ 400-550
nm), $r_{P1}$ ($\sim$ 550-700 nm), $i_{P1}$ ($\sim$ 690-820 nm),
$z_{P1}$ ($\sim$ 820-920 nm) and $y_{P1}$ ($>$ 920 nm)
\citep{ton12}. (2) The solar system survey optimized for Near Earth
Objects, which concentrates on those ecliptic directions with a wide
($w_{P1}$) band filter that roughly combines the band pass of
$g_{P1},r_{P1},i_{P1}$ filters. (3) The Medium Deep Survey, which
comprises ten fields spread across the sky and observes nightly with longer exposures in
each passband.  (4) The Stellar Transit Survey, which searches for Jupiter-like planets in close orbit around stars. (5) a Deep Survey of M31, which studies microlensing and variability in the Andromeda Galaxy.

P/2011 S1 (Gibbs) was observed on Medium-Deep field 10 (MD10),
which is centered on the DEEP2-Field 3 Multi-wavelength Survey Field,  from Aug. to Sept. 2011,
with exposure times of 565 sec to 1980 sec in the $g_{P1}$, $r_{P1}$,
$i_{P1}$, and $z_{P1}$ filters.  It was also observed in the Solar System survey from
Sep, 2010 to Nov. 2012.  It is worth noting that the first PS1
observation was in May, 2010, about one year before P/2011 S1 (Gibbs) was
discovered. 

We obtained images of P/2011 S1 (Gibbs) from the Pan-STARRS
postage stamp server. Those images were processed by PS1 image
processing pipeline (IPP) for the image detrending, astrometric
solution and photometry calibration. The ``warp" stage images have
pixel scale 0.25"/pixel or 0.2"/pixel, depending on which skycell they
are located on, and allow the better astrometric solutions for further
image stacking. The detailed observation log is shown in Table~\ref{tab1}. 
All available PS1 data are used to improve the orbital solution of P/2011 S1 (Gibbs); the result is shown in
Table~\ref{tab2}. 

\section{Cometary Activity}

To detect and trace the existence of cometary activity, we compare the radial
profile of P/2011 S1 (Gibbs) with the local PSF (Point Spread Function) at every observational epoch. 

To maximize the Signal-to-Noise Ratio (SNR) of images, all of the
usable images in each night were median combined in each filter,
centered on P/2011 S1 (Gibbs).  Another set of stacked images was
built from the same postage stamp images but centered on reference stars.
The latter were used to build the PSFs for comparison with the target radial profile and photometry calibrations.
The final reference PSF was built by averaging roughly a dozen of stars around the target
using PSF task of DAOPHOT package in IRAF\footnote{IRAF is distributed
  by the National Optical Astronomy Observatory, which is operated by
  the Association of Universities for Research in Astronomy, Inc.,
  under agreement with the National Science Foundation.}.  

\begin{figure}
\plotone{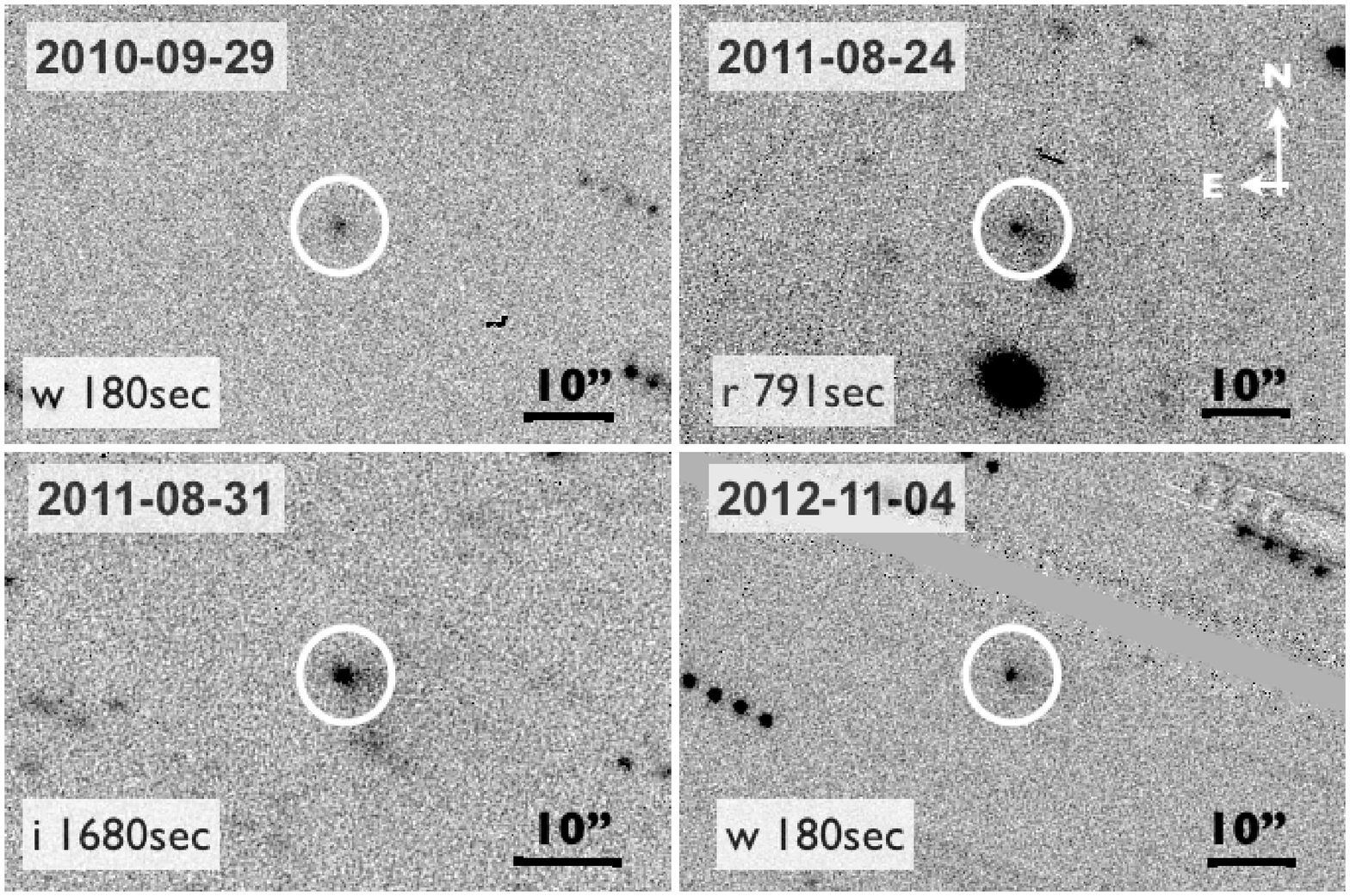}
\caption{Four P/2011 S1 (Gibbs) stacked images taken by Pan-STARRS 1. The object locates on the center of images and is marked by a white circle. First image was taken on 29 Sep., 2010, which is one year before the discovery of P/2011 S1 (Gibbs). Second (2011-08-24) and third (2011-08-31) images were obtained in the Medium Deep Survey and have longer exposure time. P/2011 S1 (Gibbs) has a clear coma in these two images. Fourth image was stacked from other set of $w_{P1}$-band images which were taken in Nov., 2012. The fuzzy shape show that P/2011 S1 (Gibbs) continue to be active at that moment. \label{fig1}}
\end{figure}

\begin{figure}
\plotone{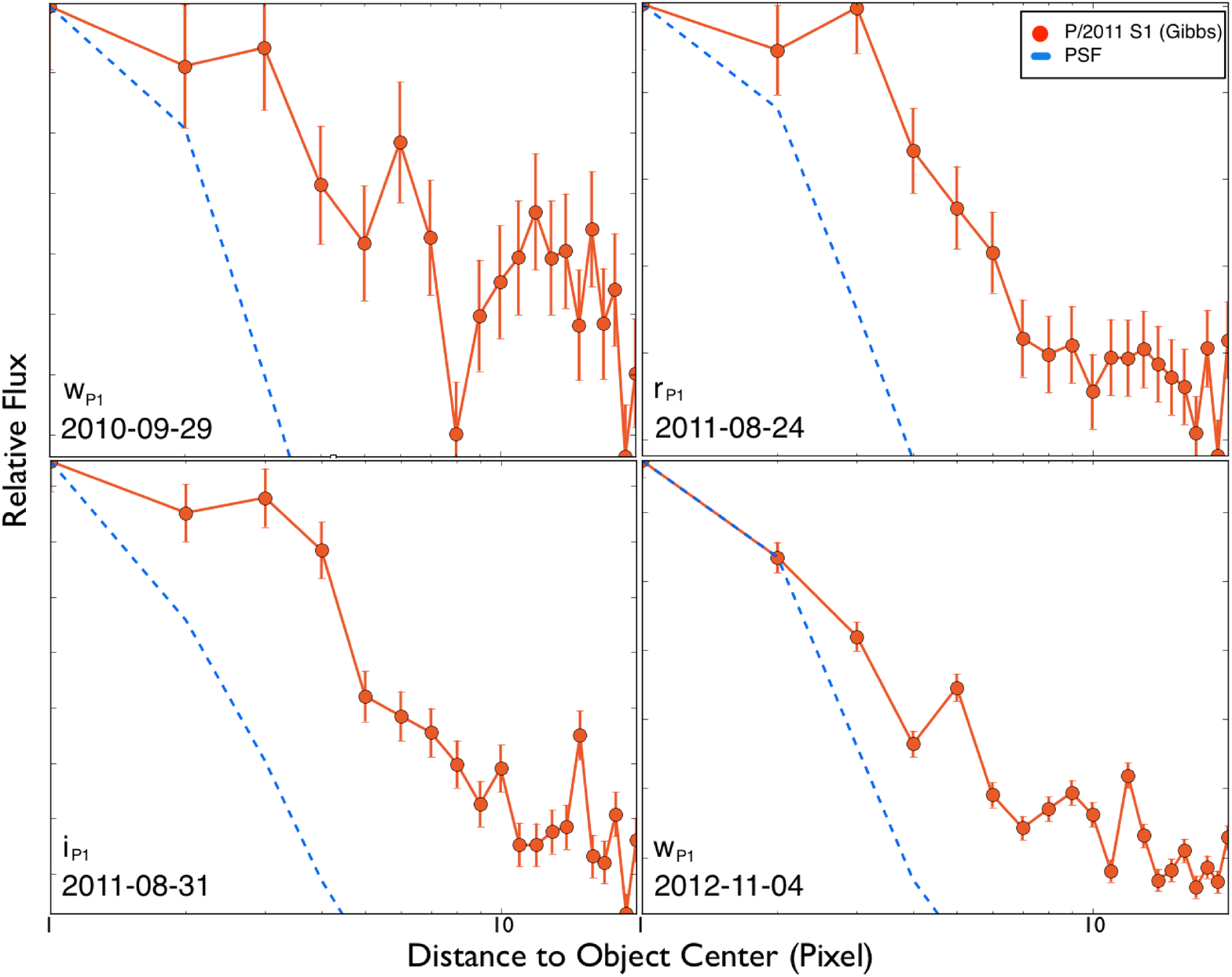}
\caption{Comparison between the P/2011 S1 (Gibbs) radial profile (red dots) and the reference PSF (dashed blue line) in the four time frames displayed in Figure 1. P/2011 S1 (Gibbs) shows a significant excess in the outer region in comparison with the PSF model, thus indicating the existence of cometary activities. \label{fig2}}
\end{figure}

Four of the P/2011 S1 (Gibbs) stacked images are shown in
Figure~\ref{fig1}.   The two $w_{P1}$-band filter images are the first
and the last observations of P/2011 S1 (Gibbs) taken by Pan-STARRS in
the solar system $w_{P1}$-band survey in 2010 and 2012.  The two other
images were taken in 2011 as part of MD10 survey.   Figure~\ref{fig2} 
shows the P/2011 S1 (Gibbs) radial profile in four different observational epochs, compared with the reference PSF, plotted with a logarithmic stretch. The radial profiles clearly show a flux excess in outer region
when compared with the stellar PSF. This suggests that P/2011 S1
(Gibbs) was continually active from 2010 to 2012. Furthermore, the
radial profile of P/2011 S1 (Gibbs) seems to change in different epochs,
hinting that the cometary activity level of this Centaur may have
some variation.  We further investigate the cometary activity of
P/2011 S1 (Gibbs) in the next section. 

\section{Photometry}
Since the P/2011 S1 (Gibbs) images were taken on different days, the
field stars are different.  Thus, we are not able to perform differential photometry.
To compare the day-to-day brightness variations, we used the PS1
absolute photometry, using the calibrated zero-point of the stacked images.

The stacked reference postage-stamp images, centered on the reference stars (see previous Section), 
were calibrated by identifying in the field PS1 catalogue stars, 
which have been calibrated with ``uber-calibration" \citep{sch12}. 
   Uber-calibration is an algorithm to photometrically calibrate
   wide-field optical imaging surveys, which was first applied on the Sloan
   Digital Sky Survey imaging data. It can simultaneously solve for
   the calibration parameters and relative stellar fluxes using
   overlapping observations \citep{pad08}.  
Those ubercaled catalogues have a relative precision (compared with
SDSS) of $<$ 10 mmag in $g_{P1}$, $r_{P1}$, and $i_{P1}$, and $\sim$
10 mmag in $z_{P1}$ and $y_{P1}$  \citep{sch12}. 
The reference images share the same zero-point with P/2011 S1 (Gibbs)
stack images which are stacked on the center of the object, so that
calibration results could be applied to those images. The stars, which
have been used to determine the zero-points in the reference images,
must satisfy the condition that there is no significant brightness
variation ($< $0.05 magnitude) in the first 2 years PS1 observations. 

To attempt to isolate the fluxes from the nucleus and the coma, multi-aperture photometry was performed using PHOT task of DAOPHOT package in IRAF.
This multi-aperture photometry does not subtract the sky background
for two principal reasons.  First, the sky background  has already
been removed in the PS1 postage stamp image; background flux is around
zero. Second, stellar crowding often prevents an accurate estimate of
the background.

We estimate the flux from the cometary nucleus by using a small
aperture, and we measure the coma flux by using
an outer annulus.  We carefully chose the optimal size of aperture and
inner/outer radius of the annulus.  If the aperture used to measure
the nucleus contribution is too large, then it will contain too much
flux from the coma. Similarly, if the inner radius of the outer
annulus is too small then a significant contribution from the nucleus
flux will be included when measuring the coma. Finally, the outer
radius of the annulus should not be allowed to extend to regions where
the SNR from the coma is too small.  

Thus, the PSF from the reference images is used to determine which should be the behavior of a point source and therefore of the nucleus only, without coma. A diameter of 1 FWHM of PSF for the aperture will
contain about half of the total flux and is large enough for estimating the
flux from the nucleus without including much coma contribution. An
annulus with inner and outer diameters of 3 and 5 times the FWHM of
PSF will only include $~9\%$ of the flux from the Moffat PSF and will be suitable for
measuring the coma flux. Finally, we decided that two times the flux
of 1 FWHM diameter aperture is representative of the nucleus flux, and
the coma flux is represented by $91\%$ of the flux from the annulus
with inner and outer diameters of 3 and 5 times the FWHM of PSF as the
coma flux. 

\subsection{Color and lightcurves}
\begin{figure}
\epsscale{0.7}
\includegraphics[angle=270, width = 1.0\textwidth]{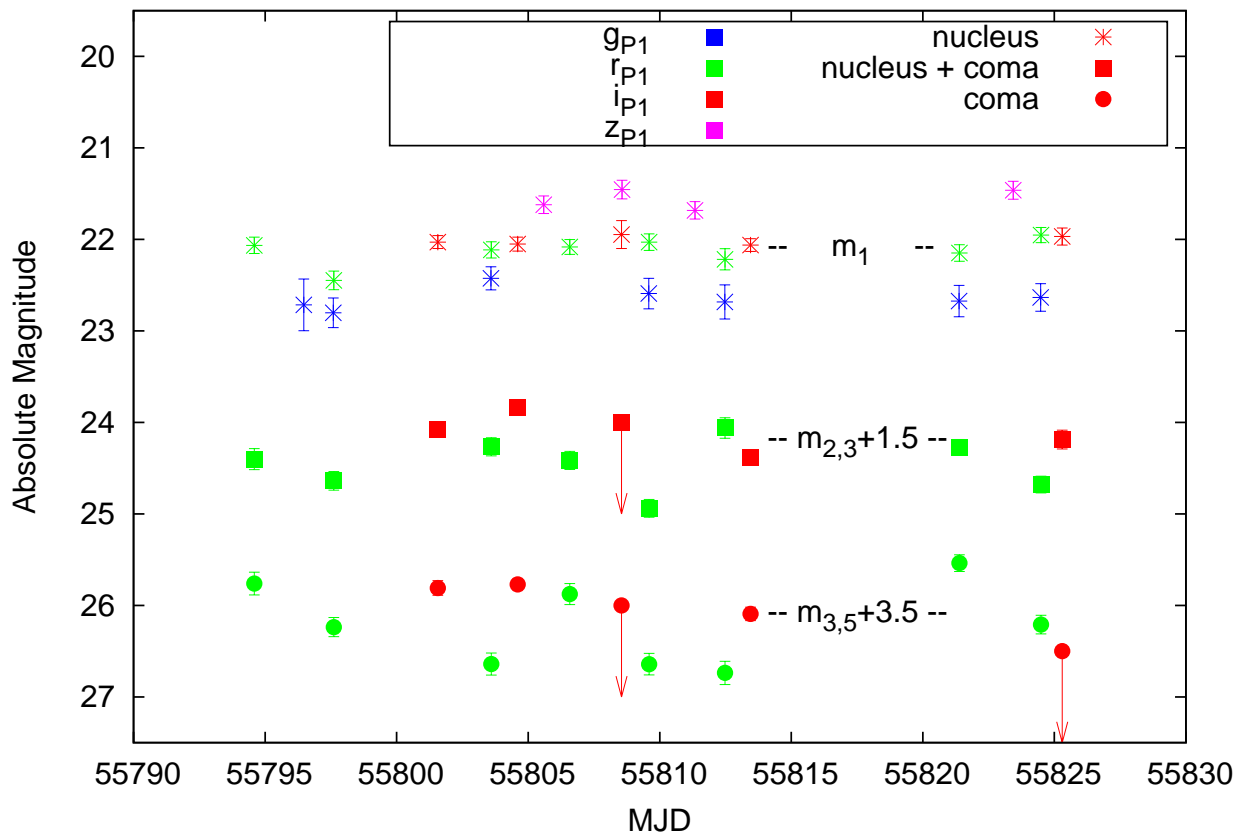}
\caption{Lightcurves of P/2011 S1 (Gibbs) in 2011. The star symbols (*) show the central region small aperture (one FWHM of PSF diameter) measurements ($m_1$), refer to the nucleus brightness. The square symbols show the intermediate region annulus (2 - 3 FWHM of PSF diameter, $m_{2,3}$) brightness, containing the flux from both coma and nucleus. The circles show the outer region annulus (3 - 5 FWHM of PSF diameter, $m_{3,5}$) after $9\%$ of the inner region flux ($9\%$ of $m_1$ flux) subtraction, are the brightness of pure coma. Some of the measurements without enough flux also show their upper limits in the plot. The lightcurves in some region ($m_{2,3}$ and $m_{3,5}$) were shifted vertically for clarity of presentation. \label{fig3}}
\end{figure}

The photometry results and day-by-day brightness variations are shown
in Figure~\ref{fig3}. The color information could be obtained only from observations acquired in 2011, given that observations were acquired in several filters, while in 2010 and 2012 only one filter ($w_{P1}$) measurements were performed, not allowing to retrieve any color information.  Only the $r_{P1}$ and $i_{P1}$ band data have
high enough SNR to permit photometry of the coma. 
The brightness variation of the coma region is significantly larger than
that of the nucleus region, at least in $r_{P1}$ filter (see Figure~\ref{fig3}). This implies that the cometary
activity is changing with time. We took the average of the measurements 
to decrease the influence of rotation. We find colors
$g_{P1}-r_{P1} = 0.52 \pm 0.06$, $r_{P1}-i_{P1} = 0.12 \pm 0.05$,
$i_{P1}-z_{P1} = 0.45 \pm 0.05$ for the nucleus region and
$r_{P1}-i_{P1} = 0.32 \pm 0.17$ for coma; the photometry results are
shown in Table~\ref{tab3}. These colors are consistent with other
active Centaurs, which are found in the blue part ($1.0 < B - R < 1.4$,
for P/2011 S1(Gibbs) is $B - R \sim 1.25$, which is calculated using the color transformation equation in \citet{ton12}) of the bimodal
color distribution of Centaurs \citep{teg08, jew09}. The coma color
seems redder than the nucleus, but it is difficult to make the
conclusion because we can not tell whether the coma is really brighter
in $r_{P1}$ band or it comes from the coma brightness variation.  

\subsection{Nucleus Size}
The brightness of the nucleus region consists of nucleus flux plus some unknown contribution of coma flux, so the flux can be used to estimate an upper limit of nucleus size.
To do so, by assuming that the nucleus has a spherical shape with a cross-section $\pi r^2$, the equivalent radius r can be calculated using the relation \citep{rus16}:

\begin{equation}
\pi r^2 p_R 10^{-0.4 \beta \alpha} = 2.25 \times 10^{22} \pi R^2 \Delta ^2 10^{-0.4(m-m_\sun)}
\end{equation}
where $p_R$ is the $R$-band geometric albedo which is almost the same as PS1 $r_{P1}$ or $i_{P1}$-band geometric albedo due to the similar band pass. $\beta$ is the linear phase coefficient of the nucleus. $\alpha$ is the solar phase angle, which together with the heliocentric distance, R, and the geocentric distance, $\Delta$, can be found in Table~\ref{tab2}.
The PS1 photometry system apparent magnitude of the Sun at Earth can be converted from standard photometry system \citep{ton12}.
The uncertainty introduced by $\beta$ is negligible when compared to that in $p_R$ (uncertain by a factor 2 or more),  if the estimation of nucleus size used the images taken near opposition ($\alpha < 3^{\circ}$).  
We assumed $\beta$ = 0.02 mag/degree \citep{mil82, mee87} and $p_R = 0.1$ \citep{kol04}.

The nucleus size estimates obtained separately from high SNR $r_{P1}$ and $i_{P1}$ band images are consistent. An upper limit to the radius of the nucleus is between 3.1 and 3.9 km.

\subsection{Mass-loss rate}
We also use the photometry to calculate the coma particle
cross-section.  Along with several assumption of coma particle properties,
we can estimate the total dust mass present within a coma-dominated
annulus. Thus, the mass loss rate can be computed by dividing the
total dust mass by the time it takes the dust to move across the
annulus. This model-dependent mass-loss rate is a good indicator to
quantify how active is P/2011 S1 (Gibbs) in comparison to other
Centaurs. 

First, we take an equivalent dust particle radius of $r_d$ = (0.1 $\mu$m $\times$ 1 cm$)^{1/2}$ $\sim$ 32 $\mu$m 
which is based on a power-law dust size of $dn/dr_d$  $\propto$ $r_d^{-3.5}$ with minimum and maximum grain radii of 0.1 $\mu$m and 1 cm \citep{jew09, li11, lac13}.
Second, the bulk density $\rho_d$ is assumed as 1000 kg/m$^3$. 
Third, the total dust cross-section within the annulus, $A_d$, can be calculated from Equation 1.
Assuming that the particle number density in coma region is very low; the column number density of particle is $<$1, then the total dust mass within the annulus is:
\begin{equation}
M_{total} = (4/3) \rho_d r_d A_d
\end{equation}
Finally, we need to assume the speed with which the dust is
crossing the coma annulus. The velocity $v_d$ is highly uncertain and
depends on the grain size \citep{cri04}. Estimates based on
macroscopic fragment ejection from 17P/Holmes \citep{ste10}, on the
coma expansion velocities of 17P/Holmes \citep{mon08, hsi10} and
C/Hale-Bopp  \citep{biv02}, and the spiral jet expansion velocity of
29P/SW1 \citep{rea13}, vary from a few 100 m/s to 1000 m/s. The
present work uses $v_d = 500$ m/s. The width of the outer region
annulus is equal to the width of 1 FWHM of PSF. Thus the
projected width is dependent on the FWHM of PSF in each image, is between
4000 km to 8000 km, and results that the dust crossing time is from 8000 second to
16000 second. 

\begin{figure}
\includegraphics[angle=270, width = 1.0\textwidth]{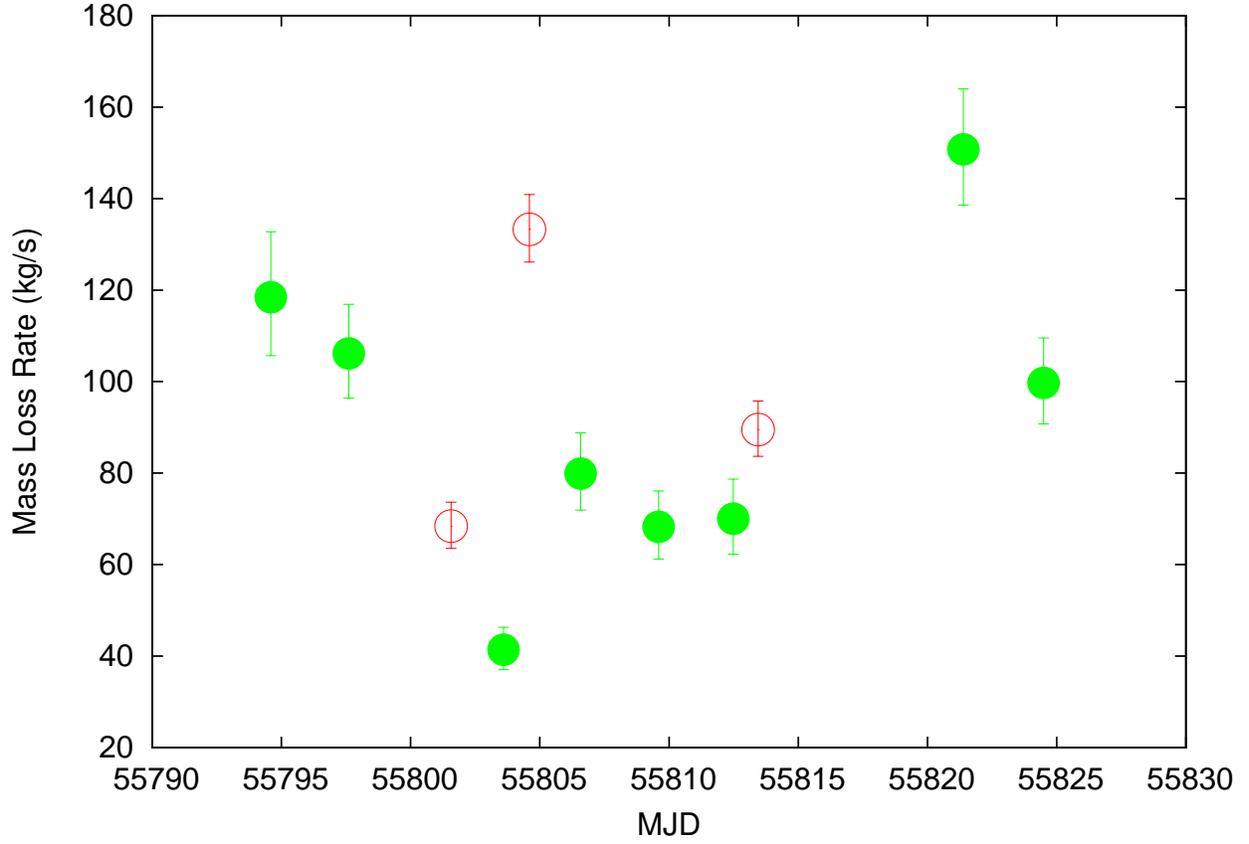}
\caption{mass-loss rate variation of P/2011 S1 (Gibbs) as a function of time. The mass-loss rates were calculated from the results of $r_{p1}$ photometry (green solid circles) and $i_{p1}$ photometry (red open circles). The uncertainty of mass-loss rates came from the photometry error. \label{fig4}}
\end{figure}

Figure~\ref{fig4} shows that the mass-loss rate changes with
time. Only $r_{P1}$ and $i_{P1}$ band images are used for the best SNR. Because the orbit is approximately circular, the change in
heliocentric distance is small, and the change of heliocentric
distance is not the major reason for the variations of mass-loss rate.
Also, the mass-loss rate varies from 40 kg s$^{-1}$ to 150 kg s$^{-1}$
suggesting that the distribution of volatility sources may not be
uniform over the surface of P/2011 S1 (Gibbs) and small-scale
outbursts could take place. Further discussions, together with possible causes for the
mass-loss, are in section 6.   

\section{Dynamical evolution}

In this section several questions are addressed: Where did P/2011 S1
(Gibbs) originate? Does it have any relation with other dynamical
classes of objects like the Jovian Trojans or Neptune Trojans? Is it
dynamically stable? How long can it remain in its current orbit? Is
its dynamical evolution similar to that of 29P/SW1 or 2060 Chiron? Or
P/2011 S1 (Gibbs), dynamically, it is a different type of object? 

To answer these questions, we performed a series of numerical orbital integrations to understand the orbital evolution of this object.
The orbital elements and covariance matrix were fitted using the $Orbfit$ code \citep{ber00}, with only data from PS1 detections.
The PS1 observations cover more than 2 years and provide better astrometric accuracy than other observations.
The resulting uncertainties are an order of magnitude smaller than those reported by JPL.
We use the N-body integration package Mercury 6.2 \citep{cha99} to
integrate the orbits of 200 massless clones plus P/2011 S1 (Gibbs) with orbital elements as obtained from PS1 observations (Table~\ref{tab2}).
The clones' orbital elements are assumed normally distributed around the PS1 solution for P/2011 S1 (Gibbs).
The calculation is stopped when the semi-major axis exceeds 1000AU. The maximum integration time is 100 million years with 8 days time step, although only very few of the clones survive the full integration.
A high-resolution integration with Bulirsch-Stoer method with 1 day time step for 5000 years is also performed to understand the dynamical evolution of the present orbit.
In all of the integrations, we only consider the gravitational force
from the Sun and planets. Non-gravitational effects i.e. out-gassing of object are omitted.

\begin{figure}
\includegraphics[angle=270, width = 1.0\textwidth]{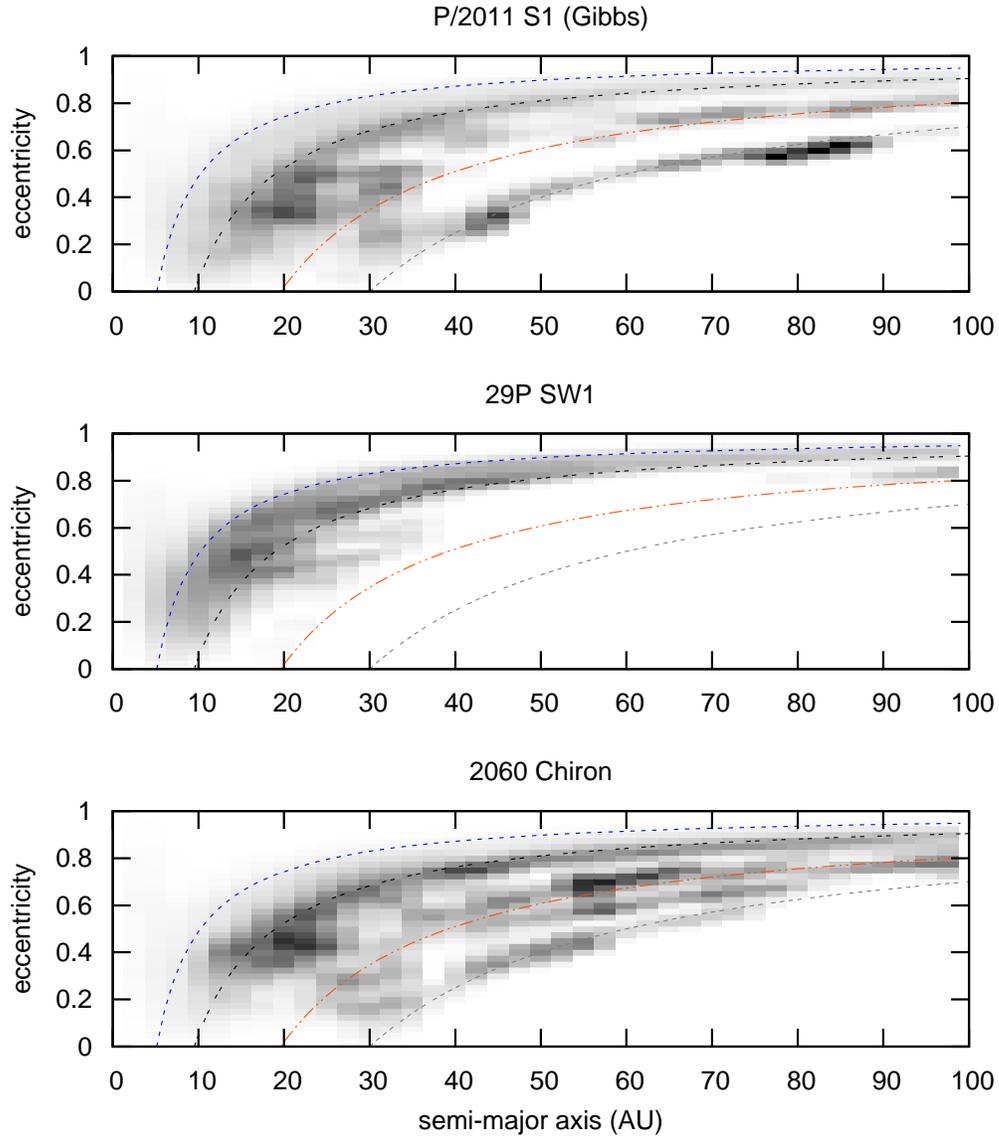}
\caption{Occupation density-map of the dynamical future of three P/2011 S1 (Gibbs), 29P/SW 1 and 2060 Chrion. Darker patches have been occupied for a longer time by clones. Dashed lines mark the perihelia of four outer planets.  \label{fig5}}
\end{figure}

The integration result shows that the orbit of P/2011 S1 (Gibbs) is
evolving chaotically; every clone has it own evolution path and is not
able to trace the precise evolution path of P/2011 S1 (Gibbs) for long
time ($<$ 1000 years). Figure~\ref{fig5} shows the dynamical evolution
plotted as an occupation density-map to present the results
statistically. 

\subsection{Dynamical similarity with 29P/Schwassmann-Wachmann 1 and (2060) Chiron}

29P/SW1 and (2060) Chiron (hereafter, 2060), two well-known active Centaurs, have different orbital elements. 29P/SW1 currently has a circular orbit without any planet crossing, but 2060 has a more eccentric orbit between 8AU and 19AU within the orbits of Jupiter and Neptune.
P/2011 S1 has a rather circular but Saturn-crossing orbit. How do the dynamical behaviors of these objects compare?
We study 200 clones for 29P/SW1 and 2060 each in the same way as P/2011 S1 (Gibbs) but with JPL orbital elements to investigate their orbital evolutions, and the results are also shown as an occupation density-map in Figure~\ref{fig5}.
Forward integration results show that  29P/SW1 is more strongly influenced by Jupiter and Saturn; most of time the perihelion of 29P/SW1 varies between the semi-major axes of Jupiter and Saturn. In all cases, 29P/SW1 is scattered into an unstable, high eccentric orbit by these two gas giants.  The dynamical lifetime is significantly shorter than the lifetimes of the other two Centaurs.
(2060) Chiron has a different dynamical trend; the perihelion shifts among the orbits of all of the four outer planets and can be scattered by any one of them.
Inspection of the occupation density-maps in Fig. 5 indicates that the dynamical evolution of P/2011 S1 (Gibbs) closely
resembles that of 2060 Chiron; P/2011 S1 (Gibbs) could also be
scattered by any of the four planets, and its dynamical lifetime is
longer than that of 29P/SW1.
Furthermore, looking at the Tisserand parameter with respect to Jupiter, $T_J$, of these objects, 29P/SW1 has a small $T_J$ (2.984), below 3, meaning that it can be classified as a member of the Jupiter-family comets, whereas, 2060 Chiron and P/2011 S1 (Gibbs) have larger $T_J$ values (3.355 for 2060, 3.122 of P/2011 S1 (Gibbs)) which explains why they are less influenced by Jupiter.


\subsection{Lifetime for current near resonance orbit with Saturn}

The high resolution integration shows that the current orbit of P/2011
S1 (Gibbs) is currently near the 6:5 orbital resonance with Saturn. It
may remain near this quasi-resonance orbit for about a thousand
years. During that time, P/2011 S1 (Gibbs) has several close
encounters with Saturn, and the subsequent orbital evolution path of
P/2011 S1 (Gibbs) becomes too chaotic to trace reliably.  

\section{Discussion}

In section 4 we attributed the variations in coma brightness of P/2011
S1 (Gibbs) to variations of its mass-loss rate and assume that the size distribution of coma dust particle remains the same. However, we cannot exclude that the changes in coma brightness are due to variations in the coma dust size distribution.  A possible scenario
that yields a sudden change in the dust particle size distribution is
as follows.  First, regular outgassing ejects mostly small particles.
Later, an outburst ejects the remaining particles that are too large to
be lifted by normal outgassing.  In this case the mass-loss rate will be larger than our estimation in section 4.

\begin{figure}
\includegraphics[angle=270, width = 1.0\textwidth]{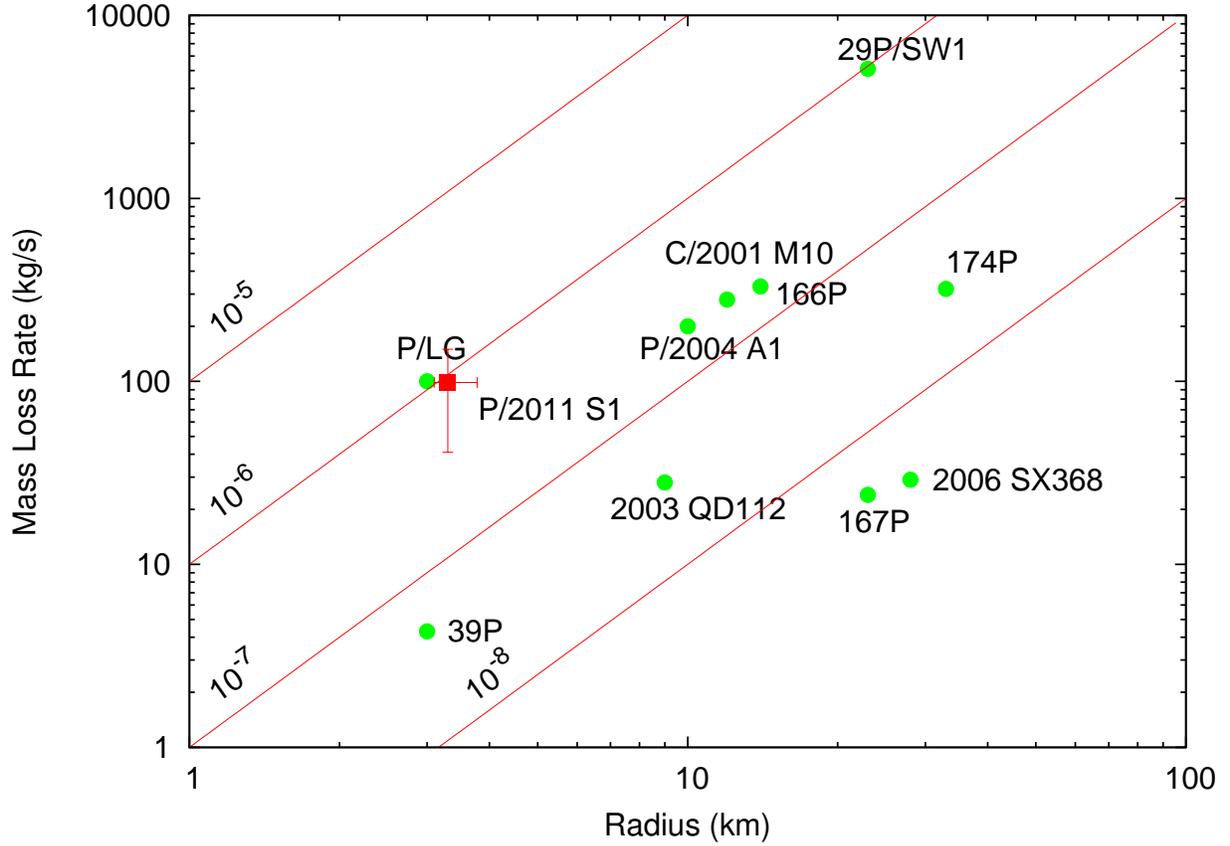}
\caption{mass-loss rate vs nucleus size. The data on P/2011 S1 (Gibbs) are from this work. P/LG is from \citet{lac13} and others from \citet{jew09}. All of the radius estimations are the upper limit due to the unknown contribution of coma. The solid line shows the specific mass-loss rate are labeled in units of $kg$ $m^{-2} s^{-1}$. Notice that except 166P and P/2011 S1 (Gibbs), other objects with mass-loss rate/Radius $> 10^{-7}$ kg/s km have JFC like Tisserand Parameter ($2 < T_J < 3$). \label{fig6}}
\end{figure}

We use PS1 photometric measurements to estimate the nucleus size and mass-loss rate of P/2011 S1 (Gibbs), and we compare the results with known active Centaurs (see Figure~\ref{fig6}). The size and mass-loss rate of 29P/SW1 and P/LG are from \citet{lac13} and data on the remaining objects are from \citet{jew09}.
To evaluate the intrinsic out-gassing activities of different objects, it is better to examine the specific mass-loss rate, i.e., the mass-loss rate per unit area.
If the mass-loss rate is normalized by the upper limit of surface area of the nucleus, a value $10^{-6}$ kg m$^{-2}$ s$^{-1}$ for P/2011 S1 (Gibbs) is obtained.
Thus, similar to 29P/SW1 and P/LG, P/2011 S1 (Gibbs) has a higher mass-loss rate per unit surface area than most other active Centaurs.
Note that except for 166P and P/2011 S1 (Gibbs), all other objects with specific mass-loss rate higher than $10^{-7}$ kg m$^{-2}$ s$^{-1}$ have $T_J$ less than 3 and can be classified as Jovian family comet. The large specific mass-loss rate of Jovian family comets is due to their smaller heliocentric distances compared with that of Centaurs.

The two major influences on the the mass-loss rate are the perihelion
distance and the composition of volatile materials.  Comparing with
other active Centaurs, P/2011 S1 (Gibbs) does not have a particularly
small perihelion. The composition of near surface volatile materials
might, thus, be the main reason for its unusually high mass-loss rate.  
29P/SW1 has been observed to display CO/CO+ emission \citep{coc82, sen94, gun02, pag13}. On the other hand, \citet{lac13} suggested that water ice is the source of activity of P/LG. For P/2011 S1 (Gibbs), the perihelion is larger than P/LG and 29P/SW1, so the water production rate should be much lower. Given its larger perihelion distance we would expect P/2011 S1 (Gibbs) to display lower specific mass-loss rate than P/LG if both are driven by water ice sublimation. For this reason, we therefore propose that the composition of P/2011 S1(Gibbs) should be similar to 29P/SW1 with CO as the major source of cometary activity. 

Moreover, the presence of significant variation of mass-loss rate of
P/2011 S1 (Gibbs), while the heliocentric distance remained the same
(See Table~\ref{tab2}), indicates that other mechanisms could affect
the outgassing rate. CO-rich ``hot spots" on the surface of this
Centaur may explain this variability. Once a hot spot is heated by the
sunlight, it can suddenly increase the mass-loss rate. Under this kind
of scheme, the time variation of mass-loss rate could be closely
related to the rotation of the nucleus, like 29P/SW1 \citep{tri08,
  tri10}. The PS1 data are not able to trace the rotation period and
out-burst period of P/2011 S1 (Gibbs), if it exists. More observations
are needed for further investigation. 

The total area of active regions can be estimated from the mass-loss
rate. Assuming the dust-to-gas mass ratio is $\sim$ 0.1 to 1
\citep{sin92, san96, kaw97}, and specific mass-loss rate of CO in 7.5
AU, 10$^{-3}$ kg m$^{-2}$s$^{-1}$ \citep{jew09}, the active region is
around 0.1\% to 1\% of the total surface area of P/2011 S1
(Gibbs). The result is also consistent with the assumption of
existence of active hot spots. 

Assuming a 100 kg s$^{-1}$ mass loss rate and 0.1\% to 1\% active surface area, we can estimate the active area recession rate of P/2011 S1 (Gibbs) to be $\sim$ 0.3 km to 3 km per thousand years.
Considering that the lifetime of P/2011 S1 (Gibbs) around current orbit is only about a thousand years, this object cannot always remain active; this activity event must be recent. A possible explanation is that the hot spots of P/2011 S1 (Gibbs) were produced by some recent impacts or other mechanisms, uncovering volatile CO ice. Once the CO hot spots have been covered by dust mantle or CO runs out, the activity of  P/2011 S1 (Gibbs) will soon stop.

The orbital integration results suggest that the future of P/2011 S1 (Gibbs) is closer to the main population of Centaurs than 29P/SW1-like objects. We are led to believe that dynamically 29P/SW1 and P/2011 S1 (Gibbs) may represent an intermediate stage between Centaurs and JFCs, with 29P/SW1 closer to the JFCs and P/2011 S1 (Gibbs) closer to Centaurs.

\section{CONCLUSIONS}
We report photometric observations of active Centaur P/2011 S1 (Gibbs), improved orbital elements obtained from PS1 survey images, and numerical simulations of its orbital evolution.
Our results can be summarized as follows:\\
(i) P/2011 S1 (Gibbs) was active in 2010, one year before the discovery by A.R Gibbs, and remained active in 2012.\\
(ii) The nucleus of P/2011 S1 (Gibbs) has a radius $< 4$ km and colors $g_{P1}-r_{P1} = 0.52 \pm 0.06$, $r_{P1}-i_{P1} = 0.12 \pm 0.05$ and $i_{P1}-z_{P1} = 0.45 \pm 0.05$, consistent with other known active Centaurs. The data also show that the coma materials appear significantly redder than the nucleus. The brightness of the coma varies with time suggesting several small-scale outburst events in the observation period. \\
(iii) The model-dependent mass-loss rate of P/2011 S1 (Gibbs) $\sim
100$ kg s$^{-1}$. The mass-loss rate per surface area is higher than
other active Centaurs and as high as 29P/Schwassmann-Wachmann 1.  It
also varies with time from $\sim 40 $ kg s$^{-1}$ to 150 kg
s$^{-1}$. This observed mass-loss rate variation is not related to the
heliocentric distance, because the orbit of P/2011 S1 (Gibbs) is rather circular. We propose the occurrence of a 29P/SW1-like outburst effect but more and long-term observations are needed to test this scenario.\\
(iv) Numerical simulations show that the future orbital evolution of P/2011 S1 (Gibbs) is more similar to that of the Centaur (2060) Chiron rather than to 29P/Schwassmann-Wachmann 1. The results also show that  P/2011 S1 (Gibbs) is dynamically unstable and can remain near its current orbit for only a thousand years or so. \\
(v) Finally, given its unusually high mass-loss rate and orbital evolution results, we have come to the conclusion that P/2011 S1 (Gibbs) has similar near-surface composition to 29P/SW1 but an orbit typical of a Centaur.

\acknowledgments

We thank for J. J. Kavelaars for his helpful comments on the manuscript.

This work was supported in part by NSC Grant: NSC 102-2119-M-008-001 and NSC 101-2119-M-008-007-MY3 and Ministry of Education under the 5500 Program NCU. 

The Pan-STARRS1 Surveys (PS1) have been made possible through contributions of the Institute for Astronomy, the University of Hawaii, the Pan-STARRS Project Office, the Max-Planck Society and its participating institutes, the Max Planck Institute for Astronomy, Heidelberg and the Max Planck Institute for Extraterrestrial Physics, Garching, The Johns Hopkins University, Durham University, the University of Edinburgh, Queen's University Belfast, the Harvard-Smithsonian Center for Astrophysics, the Las Cumbres Observatory Global Telescope Network Incorporated, the National Central University of Taiwan, the Space Telescope Science Institute, the National Aeronautics and Space Administration under Grant No. NNX08AR22G issued through the Planetary Science Division of the NASA Science Mission Directorate, the National Science Foundation under Grant No. AST-1238877, and the University of Maryland.

\clearpage

\clearpage

\clearpage

\clearpage

\clearpage

\clearpage

\clearpage

\begin{deluxetable}{lcccccc}
\tabletypesize{\scriptsize}
\tablecaption{Observation log of P/2011 S1(Gibbs) \label{tab1}}
\tablewidth{0pt}
\tablehead{\colhead{Obs Date}    & \colhead{Filter} & \colhead{$\#$ of exps} & \colhead{EXP (sec)} & \colhead{$\alpha$ (degree)} & \colhead{R (AU)} & \colhead{$\Delta$ (AU)}}
\startdata
2010-09-29     & $w_{P1}$ & 45sec $\times$ 4 & 180 & 4.1 & 7.08 &7.93 \\
2011-08-21     & $g_{P1}$ & 113sec $\times$ 5 &565 & 3.5 & 6.64 & 7.56 \\
2011-08-21     & $r_{P1}$ & 113sec $\times$ 4 &452 & 3.5 & 6.64 & 7.56 \\
2011-08-23     & $z_{P1}$ & 240sec $\times$ 6 &1440 & 3.2 &6.63 & 7.56\\
2011-08-24     & $g_{P1}$ & 113sec $\times$ 6 &678 & 3.1 & 6.62&7.56 \\
2011-08-24     & $r_{P1}$ & 113sec $\times$ 7 &791& 3.1 &6.62 & 7.56\\
2011-08-28     & $i_{P1}$ & 240sec $\times$ 6 &1440& 2.6 & 6.59&7.55 \\
2011-08-30     & $g_{P1}$ & 113sec $\times$ 7 &791& 2.3 & 6.58& 7.55\\
2011-08-30     & $r_{P1}$ & 113sec $\times$ 5 &565& 2.3 & 6.58&7.55 \\
2011-08-31     & $i_{P1}$ & 240sec $\times$ 7 &1680& 2.2 &  6.58&7.55\\
2011-09-01     & $z_{P1}$ & 240sec $\times$ 5 &1200& 2.0 & 6.57&7.55 \\
2011-09-02     & $g_{P1}$ & 113sec $\times$ 4 &452& 1.9 & 6.57&7.55\\
2011-09-02     & $r_{P1}$ & 113sec $\times$ 6 &678& 1.9 &  6.57&7.55\\
2011-09-04     & $z_{P1}$ & 240sec $\times$ 4 &960& 1.6 & 6.56 &7.55\\
2011-09-04     & $i_{P1}$ & 45sec $\times$ 2 &90& 1.6 & 6.56 &7.55\\
2011-09-05     & $g_{P1}$ & 113sec $\times$ 4 &452& 1.5  & 6.55 &7.54\\
2011-09-05     & $r_{P1}$ & 113sec $\times$ 7 &791& 1.5  & 6.55 &7.54\\
2011-09-07     & $z_{P1}$ & 240sec $\times$ 5 &1200& 1.2 & 6.55& 7.54\\
2011-09-08     & $g_{P1}$ & 113sec $\times$ 8 &904& 1.1 & 6.54& 7.54\\
2011-09-08     & $r_{P1}$ & 113sec $\times$ 5 &565& 1.1 & 6.54&7.54 \\
2011-09-09     & $i_{P1}$ & 240sec $\times$ 7 &1680& 1.0 & 6.54& 7.54\\
2011-09-17     & $g_{P1}$ & 113sec $\times$ 8 &904& 0.4 & 6.53& 7.53\\
2011-09-17     & $r_{P1}$ & 113sec $\times$ 8 &904& 0.4 & 6.53& 7.53\\
2011-09-18     & $i_{P1}$ & 240sec $\times$ 7 &1680& 0.5 & 6.53 &7.53\\
2011-09-19     & $z_{P1}$ & 240sec $\times$ 8 &1920& 0.6 & 6.53 &7.53\\
2011-09-20     & $g_{P1}$ & 113sec $\times$ 5 &565& 0.7 & 6.53 &7.53\\
2011-09-20     & $r_{P1}$ & 113sec $\times$ 6 &678& 0.7 &  6.53&7.53\\
2011-09-21     & $i_{P1}$ & 240sec $\times$ 7 &1680& 0.9 &  6.53&7.53\\
2012-10-09     & $w_{P1}$ & 45sec $\times$ 4 &180& 0.6 & 6.18& 7.17\\
2012-11-04     & $w_{P1}$ & 45sec $\times$ 4 &180& 4.2 & 6.30& 7.15\\
\enddata
\tablecomments{$\alpha$ is solar phase angle, R is heliocentric distance and $\Delta$ is geocentric distance.}
\end{deluxetable}

\begin{table}
\caption{Improved orbital elements of P/2011 S1(Gibbs) \label{tab2}}
\begin{tabular}{lccccc}
\tableline
\tableline
Property & Value\\
\tableline
Semimajor axis, $a$ & 8.6016 $\pm$ 0.0003 AU &\\
Eccentricity, $e$  & 0.199602 $\pm$ $8 \times 10^6$ &\\
Inclination, $i$  & $2.681^{\circ}$&\\
Argument of perihelion, $\omega$  & $194.062 \pm 0.008^{\circ}$& \\
Longitude of ascending node, $\Omega$  & $218.897 \pm 0.001^{\circ} $&\\
Next perihelion passage  & 2014 Sep. 5 &\\
Perihelion distance, $q$  & $6.8847 \pm 0.0004 $ AU& \\
Aphelion distance, $Q$  & $10.3185 \pm 0.0006 $ AU&\\
\tableline
\tablecomments{The orbital elements solved by $Orbfit$ code \citep{ber00} with only PS1 detections.}
\end{tabular}
\end{table}

\begin{table}
\caption{Photometry \label{tab3}}
\begin{tabular}{lccccc}
\tableline
\tableline
Region measured & g band & r band & i band & z band\\
\tableline
$m_1$ & $22.51 \pm 0.04$ & $21.99 \pm 0.05$ & $21.87 \pm 0.02$ & $21.42 \pm 0.05$\\
$m_{2,3}$ & -- & $22.96 \pm 0.09$ & $22.62 \pm 0.10$ & --\\
$m_{3,5}$ & -- & $22.71 \pm 0.15$ & $22.39 \pm 0.07$ & --\\
\tableline
\end{tabular}
\end{table}





\begin{thebibliography}{}

\bibitem[Bernstein 
\& Khushalani(2000)]{ber00} Bernstein, G., \& Khushalani, B.\ 2000, \aj, 120, 3323 

\bibitem[Biver et 
al.(2002)]{biv02} Biver, N., Bockel{\'e}e-Morvan, D., Colom, P., et al.\ 2002, Earth Moon and Planets, 90, 5 


\bibitem[Bus et al.(1991)]{bus91} Bus, S.~J., A'Hearn, M.~F., 
Schleicher, D.~G., \& Bowell, E.\ 1991, Science, 251, 774 


\bibitem[Chambers(1999)]{cha99} Chambers, J.~E.\ 1999, 
\mnras, 304, 793 


\bibitem[Cochran et al.(1982)]{coc82} Cochran, A.~L., 
Cochran, W.~D., \& Barker, E.~S.\ 1982, \apj, 254, 816 


\bibitem[Crifo et al.(2004)]{cri04} Crifo, J.~F., Fulle, M., 
K{\"o}mle, N.~I., \& Szego, K.\ 2004, Comets II, 471 

\bibitem[Emel'yanenko et al.(2005)]{eme05} Emel'yanenko, 
V.~V., Asher, D.~J., \& Bailey, M.~E.\ 2005, \mnras, 361, 1345 

\bibitem[Fernandez 
\& Gallardo(1994)]{fer94} Fernandez, J.~A., \& Gallardo, T.\ 1994, \aap, 281, 911 

\bibitem[Gibbs et al.(2011)]{gib11} Gibbs, A.~R., Tornero, 
S.~F., \& Williams, G.~V.\ 2011, \iaucirc, 9234, 1 

\bibitem[Gladman et al.(2008)]{gla08} Gladman, B., Marsden, 
B.~G., \& Vanlaerhoven, C.\ 2008, The Solar System Beyond Neptune, 43 

\bibitem[Gunnarsson et al.(2002)]{gun02} Gunnarsson, M., 
Rickman, H., Festou, M.~C., Winnberg, A., 
\& Tancredi, G.\ 2002, \icarus, 157, 309 

\bibitem[Horner et al.(2004)]{hor04} Horner, J., Evans, 
N.~W., \& Bailey, M.~E.\ 2004, \mnras, 354, 798 

\bibitem[Hsieh et al.(2010)]{hsi10} Hsieh, H.~H., 
Fitzsimmons, A., Joshi, Y., Christian, D., 
\& Pollacco, D.~L.\ 2010, \mnras, 407, 1784 


\bibitem[Jewitt(2009)]{jew09} Jewitt, D.\ 2009, \aj, 137, 
4296 

\bibitem[Kaiser et al.(2010)]{kai10} Kaiser, N., Burgett, W., 
Chambers, K., et al.\ 2010, \procspie, 7733,  

\bibitem[Kawakita et al.(1997)]{kaw97} Kawakita, H., Furusho, 
R., Fujii, M., \& Watanabe, J.-I.\ 1997, \pasj, 49, L41 

\bibitem[Kolokolova et al.(2004)]{kol04} Kolokolova, L., 
Hanner, M.~S., Levasseur-Regourd, A.-C., 
\& Gustafson, B.~{\AA}.~S.\ 2004, Comets II, 577 


\bibitem[Lacerda(2013)]{lac13} Lacerda, P.\ 2013, \mnras, 
428, 1818 


\bibitem[Levison 
\& Duncan(1997)]{lev97} Levison, H.~F., \& Duncan, M.~J.\ 1997, \icarus, 127, 13 


\bibitem[Li et al.(2011)]{li11} Li, J., Jewitt, D., Clover, 
J.~M., \& Jackson, B.~V.\ 2011, \apj, 728, 31 


\bibitem[Luu 
\& Jewitt(1990)]{luu90} Luu, J.~X., \& Jewitt, D.~C.\ 1990, \aj, 100, 913 


\bibitem[Luu et al.(2000)]{luu00} Luu, J.~X., Jewitt, D.~C., 
\& Trujillo, C.\ 2000, \apjl, 531, L151 


\bibitem[Meech 
\& Jewitt(1987)]{mee87} Meech, K.~J., \& Jewitt, D.~C.\ 1987, \aap, 187, 585 


\bibitem[Meech 
\& Belton(1990)]{mee90} Meech, K.~J., \& Belton, M.~J.~S.\ 1990, \aj, 100, 1323 


\bibitem[Millis et al.(1982)]{mil82} Millis, R.~L., Ahearn, 
M.~F., \& Thompson, D.~T.\ 1982, \aj, 87, 1310 


\bibitem[Montalto et 
al.(2008)]{mon08} Montalto, M., Riffeser, A., Hopp, U., Wilke, S., \& Carraro, G.\ 2008, \aap, 479, L45 

\bibitem[Padmanabhan et al.(2008)]{pad08} Padmanabhan, N., 
Schlegel, D.~J., Finkbeiner, D.~P., et al.\ 2008, \apj, 674, 1217 

\bibitem[Paganini et al.(2013)]{pag13} Paganini, L., Mumma, 
M.~J., Boehnhardt, H., et al.\ 2013, \apj, 766, 100 

\bibitem[Reach et al.(2013)]{rea13} Reach, W.~T., Kelley, 
M.~S., \& Vaubaillon, J.\ 2013, arXiv:1306.2381

\bibitem[Romon-Martin et 
al.(2003)]{rom03} Romon-Martin, J., Delahodde, C., Barucci, M.~A., de Bergh, C., \& Peixinho, N.\ 2003, \aap, 400, 369 


\bibitem[Russell(1916)]{rus16} Russell, H.~N.\ 1916, \apj, 
43, 173 

\bibitem[Sanzovo et 
al.(1996)]{san96} Sanzovo, G.~C., Singh, P.~D., \& Huebner, W.~F.\ 1996, \aaps, 120, 301 


\bibitem[Schlafly et al.(2012)]{sch12} Schlafly, E.~F., 
Finkbeiner, D.~P., Juri{\'c}, M., et al.\ 2012, \apj, 756, 158 


\bibitem[Senay 
\& Jewitt(1994)]{sen94} Senay, M.~C., \& Jewitt, D.\ 1994, \nat, 371, 229 

\bibitem[Singh et al.(1992)]{sin92} Singh, P.~D., de Almeida, 
A.~A., \& Huebner, W.~F.\ 1992, \aj, 104, 848 

\bibitem[Stevenson et al.(2010)]{ste10} Stevenson, R., 
Kleyna, J., \& Jewitt, D.\ 2010, \aj, 139, 2230 

\bibitem[Tegler et al.(2008)]{teg08} Tegler, S.~C., Bauer, 
J.~M., Romanishin, W., 
\& Peixinho, N.\ 2008, The Solar System Beyond Neptune, 105 

\bibitem[Tiscareno 
\& Malhotra(2003)]{tis03} Tiscareno, M.~S., \& Malhotra, R.\ 2003, \aj, 126, 3122 

\bibitem[Tonry et al.(2012)]{ton12} Tonry, J.~L., Stubbs, 
C.~W., Lykke, K.~R., et al.\ 2012, \apj, 750, 99 


\bibitem[Trigo-Rodr{\'{\i}}guez et 
al.(2008)]{tri08} Trigo-Rodr{\'{\i}}guez, J.~M., Garc{\'{\i}}a-Melendo, E., Davidsson, B.~J.~R., et al.\ 2008, \aap, 485, 599 


\bibitem[Trigo-Rodr{\'{\i}}guez et al.(2010)]{tri10} 
Trigo-Rodr{\'{\i}}guez, J.~M., Garc{\'{\i}}a-Hern{\'a}ndez, D.~A., 
S{\'a}nchez, A., et al.\ 2010, \mnras, 409, 1682 

\bibitem[Womack 
\& Stern(1997)]{wom97} Womack, M., \& Stern, S.~A.\ 1997, Lunar and Planetary Institute Science Conference Abstracts, 28, 1575 
\end{thebibliography}
\end{document}